\documentclass[journal=jpcafh,manuscript=article]{achemso}
\usepackage{amsfonts}
\usepackage{amsmath}
\usepackage{amssymb}
\usepackage{graphicx}
\usepackage{epstopdf}
\usepackage{hyperref} 
\usepackage{color}
\usepackage{bbold}
\usepackage[dvipsnames]{xcolor}
\usepackage{pifont} 
\usepackage[utf8]{inputenc} 
\usepackage[T1]{fontenc}

\renewcommand{\vec}[1]{\boldsymbol{#1}}

\newcommand{\CR}[1]{{\color{black} #1}}

\author{Xian-Peng Zhang}
\affiliation{Centre for Quantum Physics, Key Laboratory of Advanced Optoelectronic Quantum Architecture and Measurement (MOE), School of Physics, Beijing Institute of Technology, Beijing, 100081, China}

\alsoaffiliation{International Center for Quantum Materials, Beijing Institute of Technology, Zhuhai, 519000, China}

\author{Ting-Rui Hu}

\affiliation{Centre for Quantum Physics, Key Laboratory of Advanced Optoelectronic Quantum Architecture and Measurement (MOE), School of Physics, Beijing Institute of Technology, Beijing, 100081, China}

\author{Zhen-Biao Yang}
\affiliation{Fujian Key Laboratory of Quantum Information and Quantum Optics,
College of Physics and Information Engineering,
Fuzhou University, Fuzhou, Fujian 350116, China}

\author{Rubén Seoane Souto}
\affiliation{Instituto de Ciencia de Materiales de Madrid (ICMM), Consejo Superior de Investigaciones Científicas (CSIC), Sor Juana Inés de la Cruz 3, 28049 Madrid, Spain}
\affiliation{Quantum Advanced Research Center (QuARC),Consejo Superior de Investigaciones Cient.ficas (CSIC), Sor Juana In.s de la Cruz 3, 28049 Madrid, Spain}
\email{ruben.seoane@csic.es}

\author{Yugui Yao}
\affiliation{Centre for Quantum Physics, Key Laboratory of Advanced Optoelectronic Quantum Architecture and Measurement (MOE), School of Physics, Beijing Institute of Technology, Beijing, 100081, China}
\alsoaffiliation{International Center for Quantum Materials, Beijing Institute of Technology, Zhuhai, 519000, China}
\email{ygyao@bit.edu.cn}

\title{Quantum state tomography through multi-frequency Josephson effect}


\begin{document}

\begin{abstract}
Macroscopic quantum effects emerge when the coherent behavior of countless microscopic quantum states manifests collectively at a visible scale, as exemplified by the supercurrent carried by innumerable finite-momentum Cooper pairs.  By rigorously analyzing the time evolution of superconducting nanodevices including finite-momentum superconductors and quantum-dot Josephson junctions, we analytically derive the ubiquitous macroscopic supercurrent that originates from quantum interference between distinct many-body eigenstates, which we term as the dynamic interference supercurrent.  The dynamic interference supercurrent in a system of two Andreev spin qubits exhibits multi-frequency Josephson effect in the \textit{absence} of external voltage and is governed by the response of the many-body eigenfunctions ($\vert r\rangle$) to variations in the superconducting phase $\phi$, determined by the $\phi$-space Berry connection
$\mathcal{C}_{rr'}=\langle r|\partial_\phi|r'\rangle$. The macroscopic dynamic interference supercurrent, an inherent property of the superposition state itself, enables an advanced quantum entanglement measurement of the Andreev spin qubits. 
These findings open new avenues for investigating superconducting nanodevice properties from a wave-function level, with substantial implications for advancing quantum science and technology in solid-state superconducting devices, such as quantum computer, communication,  memory, and sensor.
\end{abstract}


\section{Introduction}
Superconducting nanodevices, which exhibit macroscopic supercurrent~\cite{josephson1962possible,cohen1962superconductive,ambegaokar1963tunneling}, have emerged as pivotal components in the development of quantum science and technology, such as quantum information processing~\cite{hays2018direct,janvier2015coherent,bretheau2013exciting,van2017microwave,nichele2020relating} that relies on demanding control over operation time. Prior theories of real-time superconducting transport have mainly relied on the powerful non-equilibrium Green's function (NEGF)~\cite{cuevas1996hamiltonian,yeyati1997resonant,sun2002hamiltonian,perfetto2009equilibrium,souto2017quench,souto2016andreev,zhang2020phase} - a widely used tool in the description of mesoscopic superconductivity~\cite{bergeret2018nonequilibrium,bergeret2005odd}. \CR{While NEGF provides an efficient description of dynamical correlations and transport observables, it does not \textit{directly} expose the amplitudes and relative phases of the underlying many-body states. Such information is essential for describing quantum interference between distinct many-body states and for accessing state-specific properties such as quantum coherence and entanglement.}

\begin{figure}[t!]
\begin{center}
\includegraphics[width=1.0\linewidth]{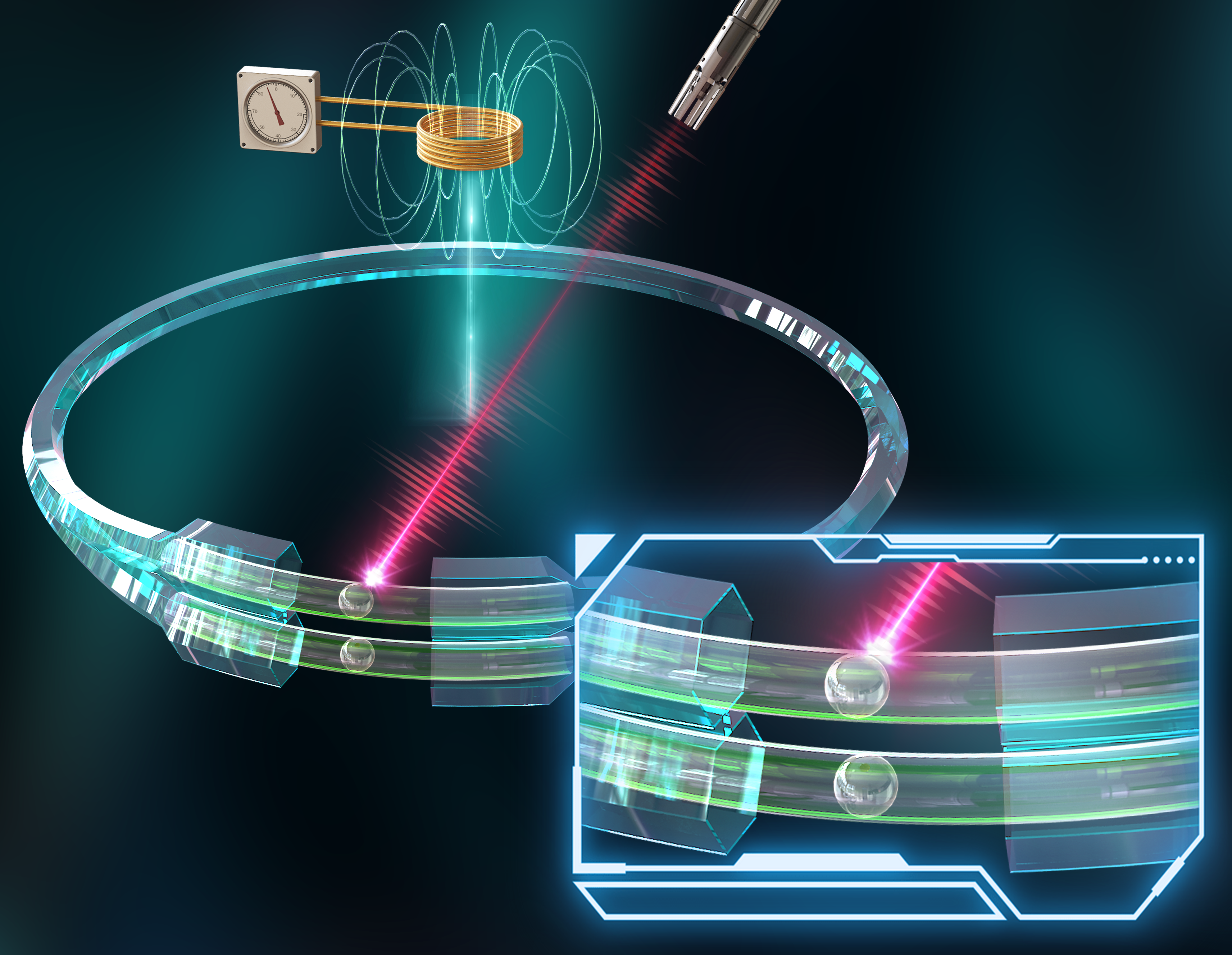}
\end{center}
\caption{Sketch of double-quantum-dot Josephson junction~\cite{kurtossy2021andreev,debbarma2023josephson,herrmann2010carbon}. Two quantum dots are gate-formed in the middle of two parallel semiconductor nanowires and are tunnel-coupled to left and right proximitized semiconductor-superconductor leads. }   
\label{SETUP} 
\end{figure}

\CR{This state-level information is particularly important in nonuniform superconductors, where the superconducting state is intrinsically many-body even in the absence of interactions when expressed in the electron picture, and its exponential form~\cite{yanagisawa2003lattice,mitake2002possible,yanagisawa2002stripes} makes its direct treatment challenging.} Consequently, i) the exact wave functions of the superconducting condensate in nonuniform superconductors, due to its intrinsic complexity, have received limited attention, mostly related to transport properties, such as the macroscopic supercurrent; ii) both static and dynamic quantum transport in superconducting systems heavily relies on the cumbersome and often unintuitive NEGF~\cite{cuevas1996hamiltonian,yeyati1997resonant,sun2002hamiltonian,perfetto2009equilibrium,souto2017quench,souto2016andreev,zhang2020phase} and density functional theory~\cite{weston2016linear,stefanucci2010time,seoane2021transient} via a numerical approach.  Recently, many-body eigenfunctions of arbitrary nonuniform superconductors were rigorously derived in both Bogoliubon and electron picture~\cite{zhang2024fermi,zhang2024fabry,zhang2024non,zhang2024renormalized}, \CR{providing a direct wave-function route to macroscopic properties of superconducting nanodevices. This development opens a state-based description of superconducting transport, conceptually analogous to the role of Berry curvature in capturing intrinsic transport from the quantum geometry of Bloch states~\cite{xiao2010berry}.} 


Here, we employ a microscopic wave-function approach to quantitatively investigate the macroscopic quantum interference originating from microscopic states. Specifically, we derive an analytic expression for the supercurrent in inhomogeneous superconductors that explicitly depends on the underlying quantum state, thereby going beyond the conventional Ginzburg–Landau framework. We test our solutions in two different scenarios, including finite-momentum superconductors and quantum-dot Josephson junctions with the later realising two Andreev spin qubits. We find a dynamic interference supercurrent (DIS) that arises from the quantum interference between different many-body eigenstates and exhibits multi-frequency  Josephson effect \textit{without} external voltage.  The simple probability amplitude and oscillation frequency dependencies of the macroscopic DIS can be used for a quantum state tomography, allowing the  entanglement measurement of Andreev qubits.

\section{Model and theory}
Our methodology involves a quadratic mean-field Hamiltonian that describes an arbitrary nonuniform superconductor. We use an overcomplete basis set that contains both particle and hole operators of  \emph{each} electron that enables exact solvability~\cite{bernevig2013topological,de2018superconductivity,tinkham2004introduction,zhu2016bogoliubov} $H=\frac{1}{2}\sum^{N}_{l=1}\sum_{\sigma=\Uparrow/\Downarrow}\sum_{\eta=+/-} E^{}_{l\sigma\eta}\gamma_{l\sigma\eta}^{\dagger} \gamma_{l\sigma\eta }^{}+\mathcal{E}$. 
Here, $N$ represents the system size, and $\sigma=\Uparrow/\Downarrow$ denotes pseudospin-up and -down quasiparticles in the presence of spin-flip effects from spin-orbit coupling and spin-dependent tunneling. The above Hamiltonian involves the Nambu and spin space~\cite{de2018superconductivity}, and an additional index $\eta=+/-$ labels the high/low energy levels of each pseudospin species ($E_{l\sigma+}>E_{l\sigma-}$). 
The Bogoliubon operator associated with energy $E_{l\sigma\eta}$ is represented as a unit vector in the Nambu-spin space
\begin{align} \label{fvdafvfd}
   \gamma_{l\sigma\eta}= \sum^{N}_{n=1}\sum_{s=\uparrow,\downarrow}\left[(u_{\eta})^{\sigma s}_{ln}c^{}_{ns}+(v_{\eta})^{\sigma s}_{ln}(-sc^{\dagger}_{n-s})\right].
\end{align}
The field operator $c_{ns}^{}$ annihilates an electron with spin $s$ and quantum number $n$.  The $N\times N$ matrices $u_{s\eta}$ and $v_{s\eta}$ describe the electron and hole distributions of all Bogoliubons, respectively. The constant energy $\mathcal{E}$ is irrelevant to the physics we concern here. Traditionally, studies of superconducting systems have focused primarily on the quasiparticle spectrum, including excitation energies, superconducting gaps, and associated low-energy excitations. The Ginzburg–Landau framework~\cite{ginzburg2009theory,pujos1959microscopic,de2018superconductivity,tinkham2004introduction}, for example, provides a highly successful description of superconductivity in terms of a macroscopic order parameter and the corresponding energy scales. However, for high-precision quantum measurement and computation, knowledge of the quasiparticle spectrum alone may be insufficient, as the superconducting state encodes substantially richer information at the wave-function level.

The rigorous time evolution of a nonuniform superconductor necessitates its precise many-body eigenfunctions. Following the logical framework of Ref.~\cite{zhang2024fermi}, we divide the overcomplete basis set into two orthonormal basis sets dual to each other --  $\{\gamma^{}_{l\sigma+}\}$ for all $l$ and $\sigma$ as well as $\{\gamma^{}_{l\sigma-}\}$ for all $l$ and $\sigma$. Defined as $\gamma_{l\sigma\eta}\vert V\rangle_{\eta}=0$ for  all $l$ and $\sigma$, the effective vacuum states (EVSs) are given by  
\begin{equation} \label{trufvdfmmvl}
    \vert V\rangle_{\eta}=\frac{1}{N^{1/2}_{\eta}}\prod^{N}_{\jmath=1}\left(1+e^{i\phi^{\eta}_\jmath}\mathcal{A}^{\eta}_{\jmath\jmath}c^{\dagger}_{\jmath\Uparrow\eta}c^{\dagger}_{\jmath\Downarrow\eta}\right)\vert 0\rangle,
\end{equation}
where $N_{\eta}=\prod_{\jmath}\left(1+\vert\mathcal{A}^{\eta}_{\jmath\jmath}\vert^2\right)$. The superconducting pseudospin cloud operators show how the electrons pair with each other in a nonuniform superconductor
\begin{align} \label{andreevcloud}
c^{\dagger}_{\jmath\sigma\eta}=\sum_{ns}c^{\dagger}_{ns}(U_{\eta})^{ s\sigma }_{n\jmath}.
\end{align}
The EVS \eqref{trufvdfmmvl} is full of  Bogoliubov-like singlet -- a superposition of vacuum state and Cooper pair. The detailed expressions of $\mathcal{A}^{\eta}_{\jmath}$, $\phi^{\eta}_{\jmath}$, and $U_{\eta}$ are given in Ref.~\cite{zhang2024fermi}.   The superconductor has $4^{N}$ many-body eigenstates, including $C(2N,n)$ $n$-Bogoliubon eigenstates, where  $C(2N,n)=(2N)!/[(2N-n)!n!]$ is  combination formula. The ground state is a filled sea of all negative quasiparticles in the orthonormal basis set $(\gamma_{l\Uparrow+},\gamma_{l\Downarrow+})$ from $\vert V\rangle_{+}$. The same goes for the $\eta=-$ orthonormal basis set $(\gamma^{}_{l\Uparrow-},\gamma^{}_{l\Downarrow-})$ with the EVS $\vert V\rangle_{-}$. Therefore,  we obtain the ground state of arbitrary nonuniform superconductor
\begin{align} \label{yfvklal}
    \vert G \rangle=\left(\prod_{E_{l\sigma+}<0}\gamma^{\dagger}_{l\sigma+}\right)\vert V\rangle_+=\left(\prod_{E_{l\sigma-}<0}\gamma^{\dagger}_{l\sigma-}\right)\vert V\rangle_-,
\end{align}
whose ground-state energy is given by 
\begin{align} \label{fdjvkdf}
    \mathcal{E}_G=\mathcal{E}_{+}+\sum_{E_{l\sigma+}<0}E_{l\sigma+}=\mathcal{E}_{-}+\sum_{E_{l\sigma-}<0}E_{l\sigma-}.
\end{align}
Here, $\mathcal{E}_{\eta}=\mathcal{E}+\sum_{\eta'\neq\eta}\frac{1}{2}E^{}_{ls\eta'}$ is  the EVS energy of the corresponding orthonormal basis set. Distinguished from other materials, the uniqueness of the nonuniform superconductor is the electrically detectable signatures of the macroscopic supercurrent - an inherent property in the flux-tunable superconducting state. Using the exact wave function of the arbitrary superconducting state, we are able to derive  macroscopic quantum effects of superconducting nanodevices directly from microscopic principles. Next, we derive the rigorous time evolution of superconducting nanodevices and analytically derive the macroscopic supercurrent of the arbitrary superposition initial state in different scenarios, including finite-momentum superconductor and quantum-dot Josephson junction.

\section{DIS in a finite-momentum superconductor}
We begin with a simple but intuitive scenario--finite-momentum superconductor with a single-$q$ order parameter $\Delta(x)=\Delta e^{iqx}$~\cite{daido2022intrinsic,kaur2005helical,yuan2022supercurrent}. The momentum of Cooper pairs, denoted by $\hbar q$, increases with the injected current before the superconducting gap closes~\cite{kaur2005helical}. The ground state, Eq.~\eqref{yfvklal}, is composed by finite-momentum Bogoliubov singlets, i.e., $\vert G \rangle=\prod_{k}[u_k(q)+v_k(q)c^{\dagger}_{k+q/2\uparrow}c^{\dagger}_{-k+q/2\downarrow}]\vert 0\rangle$, where $c^{}_{ks}$ is the field operator for an electron with momentum $\hbar k$, spin $s$, and energy $\epsilon_{k}$. The Bogoliubov coefficients $u_k(q)$ and $v_k(q)$, relying on the order parameter $\Delta$, are given by $u_{k}(q)=\frac{1}{\sqrt{2}}\{1+\epsilon^{}_{k,+}(q)/\sqrt{\Delta^2+\epsilon^{2}_{k,+}(q)}\}^{1/2}$ and $v_{k}(q)=\frac{1}{\sqrt{2}}\{1-\epsilon^{}_{k,+}(q)/\sqrt{\Delta^2+\epsilon^{2}_{k,+}(q)}\}^{1/2}$  with $\epsilon^{}_{k,\pm}(q)=\frac{1}{2}(\epsilon_{k+q/2}\pm\epsilon_{-k+q/2})$. The  ground-state energy \eqref{fdjvkdf} becomes $\mathcal{E}_G=\mathcal{E}+\frac{1}{2}\sum_{ks}E_{ks-}(q)$, where $E_{ks\eta}(q)=\epsilon^{}_{k,-}(q)+\eta\sqrt{\Delta^2+\epsilon^{2}_{k,+}(q)}$ is the Bogoliubov energy.

Now, we study the supercurrent of the finite-momentum superconductor from the wave-function perspective.  Notably, the superconducting spin cloud \eqref{andreevcloud} of the finite-momentum superconductor,  represented by $c^{\dagger}_{ks}$ for all $k$ and $s$,  possesses a well-defined group velocity, given by $V_k=\frac{1}{\hbar}\partial_k\epsilon_k$. This allows for the microscopic definition of the macroscopic supercurrent of an arbitrary superconducting state ($\vert X\rangle$) from the perspective of the superconducting spin cloud. Specifically, the \CR{total} supercurrent can be expressed as $I_{\vert X\rangle}=e \sum_{ks}V_{k}P_{ks}(X)$, where $ P_{ks}(X)=\langle X\vert c^{\dagger}_{ks}c^{}_{ks} \vert X \rangle$ represents the occupancy of the superconducting spin cloud with momentum $k$ and spin $s$ in  $\vert X\rangle$. For the ground state, i.e.,  $\vert X\rangle=\vert G \rangle$, we have $P_{k+q/2\uparrow}(G)=P_{-k+q/2\downarrow}(G)=\vert v_k(q)\vert^2$ and $I_{\vert G\rangle}=\frac{2e}{\hbar}\partial_{q}\mathcal{E}_G$, recovering 
the Ginzburg-Landau theory result. Next, we calculate the supercurrent of the excited eigenstates  with one excited quasiparticle, such as $\vert X \rangle=\gamma^{\dagger}_{k+q/2\uparrow+}\vert G\rangle$. When a positive Bogoliubon $\gamma^{\dagger}_{k+q/2\uparrow+}$ is added to the ground state, the corresponding Bogoliubov-singlet state $[u_k(q)+v_k(q)c^{\dagger}_{k+q/2\uparrow}c^{\dagger}_{-k+q/2\downarrow}]\vert 0\rangle$ becomes a single-electron state  $c^{\dagger}_{k+q/2\uparrow}\vert 0\rangle$. This leads to a change in the supercurrent by an amount of $e [V_{k+\frac{q}{2}}\vert u_k(q)\vert^2-V_{-k+\frac{q}{2}}\vert v_k(q)\vert^2]$, which is equal to $\frac{2e}{\hbar}\partial_{q}E_{k\uparrow+}$. When $\vert X \rangle$ corresponds to the eigenstates, i.e., $\vert X \rangle=\vert r \rangle$, where $H\vert r\rangle=\mathcal{E}_r\vert r\rangle$,  supercurrent can be equivalently expressed by the $q$-derivative of the eigenenergy, i.e., $I_{\vert r\rangle}=\frac{2e}{\hbar}\partial_{q}\mathcal{E}_r$.  It is important to note that the current arising from the single-electron state $c^{\dagger}_{k+q/2\uparrow}\vert 0\rangle$ also contributes to the supercurrent derived from the $q$-derivative of the eigenenergy.
Thus, the prefactor $2e$ in the relation between the supercurrent and energy guarantees its equivalence to the supercurrent derived from the wave function at a microscopic level. Therefore, the macroscopic supercurrent depends on the state of the superconductor, i.e., the coherent contribution of countless superconducting pseudospin clouds (i.e., Cooper pairs) and single electrons -- carrying finite momentum.

Next, we derive the supercurrent of arbitrary superposition of eigenstates $\vert X\rangle =\sum_{r} p_{r}(t)\vert r \rangle $. It inevitably evolves with a time-dependent probability amplitude for each eigenstate, i.e., $p_{r}(t)=p_{r}(0)e^{-i\mathcal{E}_{r}t/\hbar}$. The normalization condition requires   $\sum_{r}\vert p_{r}(t)\vert^2=1$ and the state energy is $\mathcal{E}_X=\sum_{r} \vert p_{r}(t)\vert^2 \mathcal{E}_{r}$.   The occupancy of the superconducting spin cloud can be decomposed into two parts: $ P_{ks}(X)=\sum_{r} \vert p_{r}(t)\vert^2 P^{rr}_{ks}+\sum_{r\neq r'}p^{*}_{r}(t)p_{r'}(t)P^{rr'}_{ks}$, where $P^{rr'}_{ks}=\langle \mathcal{E}_{r} \vert c^{\dagger}_{ks}c^{}_{ks} \vert \mathcal{E}_{r'}\rangle $. The total supercurrent, beyond 
the Ginzburg-Landau theory~\cite{ginzburg2009theory,pujos1959microscopic,de2018superconductivity,tinkham2004introduction}, consists of a static (time-independent) supercurrent defined as the $q$ derivative of the state energy as well as a dynamic (time-dependent) supercurrent due to the quantum interference between different many-body eigenstates, i.e., $I^{}_{\vert X\rangle}=\bar{I}^{}_{\vert X\rangle}+\delta I^{}_{\vert X\rangle}$ 
with  
\begin{align} \label{fvfavmfa}
    \bar{I}_{\vert X\rangle}=\frac{2e}{\hbar}\partial_{q}\mathcal{E}_X,
\end{align} 
\begin{align} \label{fvafnkfv}
    \delta I_{\vert X\rangle}=e \sum_{r\neq r'}p^{*}_{r}(0)p_{r'}(0)e^{i(\mathcal{E}_{r}-\mathcal{E}_{r'})t/\hbar}\sum_{ks}V_{k}P^{rr'}_{ks}.
\end{align}
As the central result,
Eq.~\eqref{fvfavmfa} only contains information about the energy of the state, i.e. the modulus of $p_r$, while Eq.~\eqref{fvafnkfv} also contains information about the relative phases between states.
For example, let us consider a superposition state $\vert X\rangle =p_1(t)\vert G \rangle + p_2(t)\gamma^{\dagger}_{k+q/2\uparrow+}\gamma^{\dagger}_{-k+q/2\downarrow+}\vert G \rangle$, where $P^{1,2}_{k+q/2\uparrow}=P^{2,1}_{k+q/2\uparrow}=P^{1,2}_{-k+q/2\downarrow}=P^{2,1}_{-k+q/2\downarrow}=u_k(q)v_k(q)$ and $P^{rr'}_{ks}=0$ for all other combinations ($r\neq r'$). In this case, we obtain the DIS $\delta I_{\vert X\rangle}=-e (V_{k+q/2}+V_{-k+q/2})u_k(q)v_k(q)[p^{*}_1(t)p^{}_2(t)+p^{}_1(t)p^{*}_2(t)]$, which reaches its maximum when $p_1(t)=e^{\pm i\pi} p_2(t)$ and becomes zero when $p_1(t)=e^{\pm i\pi/2}p_2(t)$. This DIS should be ubiquitous in different nonuniform superconducting system.  


\begin{figure}[t!]
\begin{center}
\includegraphics[width=1.0\linewidth]{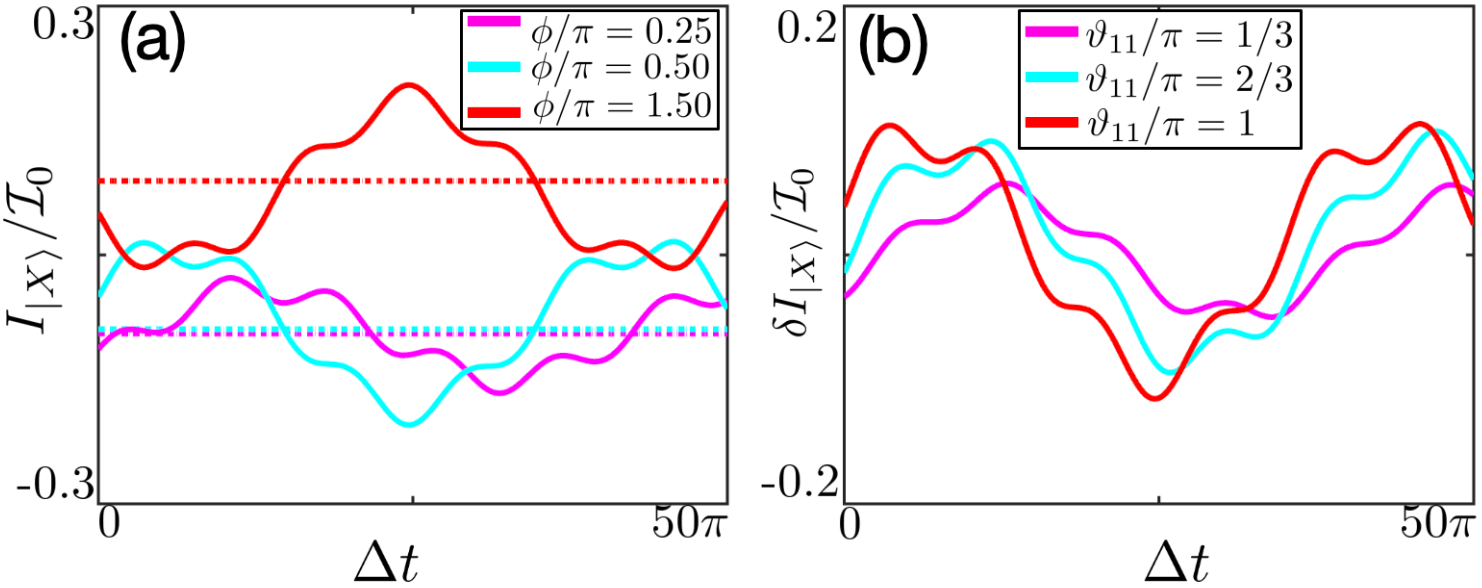}
\end{center}
\caption{The time evolution of (a) total supercurrent and (b) DIS, where we study a superposition of four many-body  eigenstates $\vert X\rangle =p_{00}\vert 0 0\rangle + p_{01}\vert 0 0\rangle+p_{10}\vert 1 0\rangle+p_{11}\vert 11\rangle$ with $p_r=\vert p_r\vert e^{i\vartheta_r}$. The dashed lines in panel (a) correspond to the static supercurrent \eqref{fvfavmfj}. The phase of probability amplitude influences the oscillation amplitude of the DIS \eqref{fvaakfvk}, as observed in panel (b). Other parameters are the same as Fig.~1 in SM.}   
\label{EXP} 
\end{figure}

\section{DIS in a quantum-dot Josephson junction} 
We next study a parallel double quantum dot-superconducting lead heterostructure~\cite{kurtossy2021andreev,vaitiekenas2021zero,liu2019semiconductor,wendin2021coherent,wang2022singlet}, sketched in Fig.~\ref{SETUP}.  The quantum dots are gate-tunable with dot energy $\epsilon^n_D$ ($n=u,d$), and the order parameters of the two superconducting leads, $\Delta e^{i\phi_j}$ ($j=L,R$), have a flux-tunable phase difference $\phi = \phi_L - \phi_R$, and $h^n_D$ ($h_L$) are the Zeeman energy in quantum dot (superconducting leads)~\footnote{In our   superconducting leads (e.g. proximitized InAs/Al leads),  we assume 
open boundary conditions  with semiconducting energy spectrum $\varepsilon_j(k) = -2t_0+\mu_S + 2t_0\cos(ka)$, where we choose the chemical potential $\mu_S=-2.8\Delta$ and nearest-neighbor tunneling $t_0 =-10\Delta$, such that Fermi kinetic momentum $k_F$, defined by $\varepsilon(k_F) =0$, is close to the bottom of the band, e.g., $k_F\simeq \pi/5a$ for a discretization step $a=5$ nm}.  The dot-lead tunnel coupling strength is parameterized by $\Gamma$, which is chosen to be the same for left and right leads. The spin-orbit interaction in the dot-lead tunneling process induces a spin-flip effects~\cite{kloeffel2011strong,froning2021strong,fasth2007direct}, which is captured by  spin rotation $\mathcal{U}_{jn}=e^{i\theta_{jn}\vec{u}_{jn}\cdot\vec{s}/2}$ (see detailed Hamiltonian in Ref.~\cite{spethmann2022coupled}) and is a key ingredient for interference matrix element [see S1 in~Supporting Information (SI)~\cite{SM}].

We derive the supercurrent of an arbitraty superposition state of a quantum-dot Josephson junction from the wave-function perspective. The breaking of translational symmetry in our hybrid structure disables superconducting spin cloud to have a well-defined group velocity. The Josephson current of the arbitrary superposition state ($\vert X\rangle$) is defined as the average value of the rate of change of the electron number in the left or right lead $I^{}_{\vert X\rangle}=+e \langle X\vert \frac{d}{dt}N_L(t)\vert X\rangle=-e \langle X\vert \frac{d}{dt}N_R(t)\vert X\rangle $~\cite{cohen1962superconductive,ambegaokar1963tunneling}, with $N_{L/R}(t)=\sum_{ns}c^{\dagger}_{jns}(t)c^{}_{jns}(t)$~\footnote{This microscopic definition describes the Josephson current as the generic current that leaks out of the left lead into the right lead with no applied voltage.}.
A direct derivation of interference terms, i.e.,  $\partial_t\langle r\vert N_j(t)\vert r'\rangle$ ($r'\neq r$), incorporates the many-body correlation functions of electrons and/or quasiparticles and requires large-scale numerical calculations. Alternative derivation is conducted by $\phi$-derivative of the eigenfunctions~\cite{zhang2024non}, which becomes feasible because all many-body eigenfunctions of arbitrary nonuniform superconductors have been rigorously derived in either Bogoliubon or electron picture~\cite{zhang2024fermi}.  Different from the previous formula in Refs.~\cite{beenakker2005superconducting,anderson1964many,bagwell1992suppression,riedel1998low,beenakker1991universal,beenakker2013fermion,zhang2024fabry}, the total supercurrent consists of a static supercurrent defined as the $\phi$ derivative of the state energy 
\begin{align} \label{fvfavmfj}
    \bar{I}_{\vert X\rangle}=\frac{2e}{\hbar}\partial_{\phi}\mathcal{E}_X,
\end{align}
as well as a dynamic supercurrent due to the quantum interference of different flux-tunable eigenstates~\cite{zhang2024non} 
\begin{align} \label{fvaakfvk}
     \delta I^{}_{\vert X\rangle}&=-\frac{2e}{\hbar}\sum_{r\neq r'}p^{*}_r(0)p^{}_{r'}(0)  e^{+i(\mathcal{E}_r-\mathcal{E}_{r'})t/\hbar} (\mathcal{E}_r-\mathcal{E}_{r'})\mathcal{C}_{rr'}.
\end{align}
When $\vert X \rangle=\vert r \rangle$, the DIS vanishes and  the total supercurrent reduces to static supercurrent, i.e., $I_{\vert r\rangle}=\frac{2e}{\hbar} \frac{\partial}{\partial \phi} \mathcal{E}_{r}$. This relation still holds when considering $\phi$-dependent magnitude of order parameter~\cite{zhang2024fermi} and a canonical ensemble at non-zero temperature~\cite{zhang2024fabry}. 
\CR{The DIS \eqref{fvaakfvk} is governed by the $\phi$-modulation of the many-body eigenfunctions, encoded in the $\phi$-space Berry connection $ \mathcal{C}_{rr'}=\langle r \vert  \partial_ {\phi}\vert r'\rangle$, analogous to the momentum-space Berry connection that plays a central role in intrinsic transport in topological systems~\cite{xiao2010berry}. Its appearance underscores the crucial role of the exact superconducting wave functions in generating macroscopic quantum effects (see more details in S1 in~SI~\cite{SM}).} Notably, the ubiquitous DIS \eqref{fvaakfvk} undergoes a time oscillation with multiply frequencies \CR{determined by the energy difference between states} [Fig.~\ref{EXP}(c)].  The multi-frequency Josephson effect here occurs in the \textit{absence} of an external voltage and  is originated by the macroscopic quantum interference.  Our formalism allows us to include the effects from initial correlations in the system. This is an important extension to previous works that studied a similar problem, starting from decoupled systems and, therefore, neglected initial correlations~\cite{souto2017quench,souto2016andreev,Taranko2018Buildup,Taranko2019Transient,Wres2021Quench}. Also, in our system, the population of the Andreev bound states do not change with time. The developed DIS expression, Eq.~\eqref{fvaakfvk}, not only highlights the significant implications of the precise many-body eigenfunctions which was paid little attention before  but also offers potential for quantum state tomography, as we demonstrate in the following.

\section{Quantum state  tomography via DIS}
Quantum state tomography, capable of reconstructing full information without destroying quantum states, is one of the most significant techniques in quantum information and computation. \CR{Next, we introduce the  applications of our DIS in quantum tomography of the  Andreev qubits.} 
Let us begin with two Andreev spin qubits. We focus on the low-energy space of the hybrid system, dominated by four flux-tunable Andreev levels [see Fig.~1(a) in S2 of~SI~\cite{SM}]. The  frequencies of two Andreev spin qubits, given by $\hbar\omega_1=E_{1\Uparrow+}-E_{1\Downarrow+}$ and $\hbar\omega_2=E_{2\Uparrow+}-E_{2\Downarrow+}$, respectively, are gate-, flux-, and field-tunable. The two Andreev spin qubits acquire different frequencies by tuning the superconducting phase difference $\phi$.  Four logic states, defined from the EVS $\vert V \rangle_+$ that has a fermionic even parity, are represented by four many-body eigenstates $\vert 00\rangle=\gamma^{\dagger}_{1\Downarrow+}\gamma^{\dagger}_{2\Downarrow+}\vert V \rangle_+$, $\vert 01\rangle=\gamma^{\dagger}_{1\Downarrow+}\gamma^{\dagger}_{2\Uparrow+}\vert V \rangle_+$, $\vert 10\rangle=\gamma^{\dagger}_{1\Uparrow+}\gamma^{\dagger}_{2\Downarrow+}\vert V \rangle_+$, and $\vert 11\rangle=\gamma^{\dagger}_{1\Uparrow+}\gamma^{\dagger}_{2\Uparrow+}\vert V \rangle_+$.  Microwave photons allow to prepare an arbitrary superposition of four many-body eigenstates $\vert X\rangle =p_{00}\vert 0 0\rangle + p_{01}\vert 0 0\rangle+p_{10}\vert 1 0\rangle+p_{11}\vert 11\rangle$ with $p_r=\vert p_r\vert e^{i\vartheta_r}$, in a hybrid quantum electrodynamics architecture \cite{hays2018direct,janvier2015coherent,bretheau2013exciting,van2017microwave,nichele2020relating}.

Let us measure the modulus ($\vert p_r\vert$) of four probability amplitudes according to the static supercurrent \eqref{fvfavmfj}. Note that four probability amplitudes $p_r$ are independent of system parameters such as $\epsilon^n_D$, $\phi$, $h^n_D$, and $h^j_L$. We can determine $\vert p_r\vert$ by measuring the static supercurrent $\bar{I}^{a}_{\vert X\rangle}$ of three different values of $\phi$ which are tuned via the adiabatic process. \CR{The three steady-state current measurements $\sum_{r}\vert p^{}_{r}\vert^2 I^{a}_{\vert r\rangle}=\bar{I}^{a}_{\vert X\rangle}$, where $a=1,2,3$, together with the normalization condition $\sum_{r}\vert p^{}_{r}\vert^2=1$ lead to a complete set of equations from where the probabilities $\vert p_r\vert$ can be determined (see generalization of $N$ Andreev qubits in S3 of~SI~\cite{SM}).} 
This system of equations simply determines the modulus  of probability amplitudes of four logic states.   Next, we determine the phase ($\vartheta_r$) of the four probability amplitudes according to the DIS \eqref{fvaakfvk}. 
In S1 of~SI~\cite{SM}, we derive interference connection $\mathcal{C}_{\imath 0,\imath 1}= \mathcal{C}^{\Uparrow\Downarrow}_{2+,2+}$ and $\mathcal{C}_{0\imath,1\imath}=\mathcal{C}^{\Uparrow\Downarrow}_{1+,1+}$ with $\mathcal{C}^{\Uparrow\Downarrow}_{l+,l+}=\sum^{}_{ns}[ (u^{}_{+})^{\Downarrow s}_{ln}\partial_{\phi}(u^{*}_{+})^{\Uparrow s}_{ln}+(v_{+})^{\Downarrow s}_{ln}\partial_{\phi}(v^{*}_{+})^{\Uparrow s}_{ln}$, where we have used the Bogoliubov transformation \eqref{fvdafvfd}. This implies that the spin/rotation flip effect from the spin-orbit interaction in the dot-lead tunneling process is a key ingredient for interference connection. Notably, $\mathcal{C}_{00,11}=0$ ($\mathcal{C}_{01,10}=0$) reveals no interference between logic states $\vert 00\rangle$ ($\vert 01\rangle $) and $\vert 11\rangle$ ($\vert 10\rangle $).   
Hence, the DIS \eqref{fvaakfvk}, excluding the modes of oscillation frequencies $\omega_1+\omega_2$ and $\omega_1-\omega_2$, oscillates with the frequencies of two Andreev level qubits ($\omega_1$ and $\omega_2$), i.e.,  $\delta I_{\vert X\rangle}=\delta I^1_{\vert X\rangle}+\delta I^2_{\vert X\rangle}$ with $ \delta I^n_{\vert X\rangle}/\mathcal{I}_0=-2\hbar \omega_n\vert \mathcal{C}^{\Uparrow\Downarrow}_{n+,n+} \vert \vert P_n\vert\cos(\omega_nt-\vartheta_n-\varphi_n)$. Here, we have parameterized the phase of $\mathcal{C}^{\Uparrow\Downarrow}_{n+,n+}$ 
with $\vartheta_n$, i.e., $\mathcal{C}^{\Uparrow\Downarrow}_{n+,n+}=\vert\mathcal{C}^{\Uparrow\Downarrow}_{n+,n+}\vert e^{i\vartheta_n}$. The probability amplitude dependence of the interference current are combined into $\vert p^{}_{00}\vert \vert p_{10}\vert e^{i\vartheta_{10}}+\vert p^{}_{01}\vert \vert p_{11}\vert e^{i(\vartheta_{11}-\vartheta_{01})}=\vert P_1\vert e^{i\varphi_1}$ and $\vert p^{}_{00}\vert \vert p_{01}\vert e^{i\vartheta_{01}}+\vert p^{}_{10}\vert \vert p_{11}\vert e^{i(\vartheta_{11}-\vartheta_{10})}=\vert P_2\vert e^{i\varphi_2}$. The phase ($\vartheta_r$) can be extracted from the above system of complex equations for $\vert P_n\vert e^{i\varphi_n}$, whose values can be obtained from the experimental measurements of the DIS.  The time evolution of $I_{\vert X\rangle }$ and $\delta I_{\vert X\rangle }$ are plotted in Figs.~\ref{EXP} (a) and (b), respectively.  The phase ($\vartheta_r$) also influences the amplitude of the DIS [Fig.~\ref{EXP}(b)]. In realistic experiments, the time-dependent supercurrent can be detected using a superconducting quantum interference device magnetometer (Fig.~\ref{SETUP}), with its pickup coil positioned well above the Andreev qubit loop~\cite{zhang2024non} (see more details in Ref.~\cite{POOLE201433,fagaly2006superconducting,braginsky1995quantum}). Consequently, we achieve an electric measurement of the entanglement between Andreev spin qubits via the inherent supercurrent inside the superposition state itself.

While the supercurrent enables nondestructive measurement of Andreev qubits, the phase $\theta_r$ of the logic state $\vert r\rangle$ can be annihilated during the DIS measurement when the probability amplitudes of the logic states interfering with $\vert r\rangle$ are all zero. This scenario occurs with the four maximally entangled two-qubit Bell states: $\vert \Phi^{\pm}\rangle=\frac{1}{\sqrt{2}}\left(\vert 00\rangle\pm\vert 11\rangle\right)$ and $ \vert \Psi^{\pm}\rangle=\frac{1}{\sqrt{2}}\left(\vert 01\rangle\pm\vert 10\rangle\right)$. 
The vanishing oscillatory DIS signal ($P_n=0$) prevents the direct measurement of the phase for these four Bell states. To address this limitation, advanced quantum state tomography techniques capable of reconstructing the full density matrix information would be required~\cite{song2017continuous,ning2019deterministic,xu2021demonstration,yang2021experimental}.

\section{Conclusion}
We analytically derive the macroscopic supercurrents of arbitrary quantum superposition states that  capture a ubiquitous transient supercurrent due to the quantum interference of different many-body eigenstates. The simple probability amplitude dependence of the DIS unveils an advanced quantum tomography technique, enabling an electric measurement of quantum entanglement between two Andreev qubits. The key innovation of this technique is its capability to directly transform the entanglement information among Andreev qubits into electrical signals in the macroscopic supercurrents.  Therefore, our results show the intimate relation between the microscopic quantum state and a macroscopic quantum effect.

\begin{suppinfo}
Calculation of interference connection, quantitative study of macroscopic supercurrent in a quantum-dot Josephson junction, and generalization of $N$ Andreev qubits. 
\end{suppinfo}

\section{Author Information}
\textbf{Corresponding Author} \\
 $^*$Email: ruben.seoane@csic.es \\
 $^*$Email: ygyao@bit.edu.cn \\
 \textbf{Notes}\\
The authors declare no competing financial interest.

\section{Acknowledgment}
This work is supported by the National Natural Science Foundation of China (Grant Nos. 12505012, 12321004, W2511003, 12234003,  12475015), the China's National Key R$\&$D Program (Grant Nos. 2025YFA1411200), the Spanish Comunidad de Madrid (CM) "Talento Program" (Project 2022-T1/IND-24070), and the Spanish Ministry of Science, Innovation, and Universities (Grants Nos. PID2022-140552NA-I00, PID2025-169814NB-I00, and CEX2024-001445-S).

\providecommand{\latin}[1]{#1}
\makeatletter
\providecommand{\doi}
  {\begingroup\let\do\@makeother\dospecials
  \catcode`\{=1 \catcode`\}=2 \doi@aux}
\providecommand{\doi@aux}[1]{\endgroup\texttt{#1}}
\makeatother
\providecommand*\mcitethebibliography{\thebibliography}
\csname @ifundefined\endcsname{endmcitethebibliography}
  {\let\endmcitethebibliography\endthebibliography}{}

\end{document}